\documentclass[10pt, conference, letterpaper]{IEEEtran}

\usepackage{algorithm}
\usepackage{algorithmicx}
\usepackage{algpseudocode}
\usepackage{amsfonts}
\usepackage{amsmath}
\usepackage{amssymb}
\usepackage[utf8]{inputenc}
\usepackage{xcolor}
\usepackage{mathtools}
\usepackage{graphicx}
\usepackage{caption}
\usepackage{subcaption}
\usepackage{import}
\usepackage{multirow}
\usepackage{cite}
\usepackage[export]{adjustbox}
\usepackage{breqn}
\usepackage{mathrsfs}
\usepackage{acronym}
\usepackage{setspace}
\usepackage{bm}
\usepackage{stackengine}
\usepackage{booktabs}
\usepackage{listings}
\usepackage[acronyms,nonumberlist,nopostdot,nomain,nogroupskip]{glossaries}
\usepackage{hyperref}
\usepackage{textcomp}
\usepackage{tikz}

\lstdefinelanguage{BibTeX}
  {keywords={%
      @article,@book,@collectedbook,@conference,@electronic,@ieeetranbstctl,%
      @inbook,@incollectedbook,@incollection,@injournal,@inproceedings,%
      @manual,@mastersthesis,@misc,@patent,@periodical,@phdthesis,@preamble,%
      @proceedings,@standard,@string,@techreport,@unpublished%
      },
   comment=[l][\itshape]{@comment},
   sensitive=false,
  }

\usepackage[font=scriptsize]{subcaption}
\usepackage[font=footnotesize]{caption}

\usepackage{listings}

\newacronym{3gpp}{3GPP}{3rd Generation Partnership Project}
\newacronym{adc}{ADC}{Analog to Digital Converter}
\newacronym{5g}{5G}{5th generation}
\newacronym{aimd}{AIMD}{Additive Increase Multiplicative Decrease}
\newacronym{am}{AM}{Acknowledged Mode}
\newacronym{amc}{AMC}{Adaptive Modulation and Coding}
\newacronym{aqm}{AQM}{Active Queue Management}
\newacronym{awgn}{AGWN}{Additive White Gaussian Noise}
\newacronym{balia}{BALIA}{Balanced Link Adaptation}
\newacronym{bdp}{BDP}{Bandwidth-Delay Product}
\newacronym{bf}{BF}{beamforming}
\newacronym{cc}{CC}{Congestion Control}
\newacronym{cdf}{CDF}{Cumulative Distribution Function}
\newacronym{cn}{CN}{Core Network}
\newacronym{cqi}{CQI}{Channel Quality Information}
\newacronym{cp}{CP}{Control Plane}
\newacronym{csirs}{CSI-RS}{Channel State Information - Reference Signal}
\newacronym{dc}{DC}{Dual Connectivity}
\newacronym{rb}{RB}{Resource Block}
\newacronym{dce}{DCE}{Direct Code Execution}
\newacronym{dci}{DCI}{Downlink Control Information}
\newacronym{udp}{UDP}{User Datagram Protocol}
\newacronym{dl}{DL}{Downlink}
\newacronym{dmr}{DMR}{Deadline Miss Ratio}
\newacronym{dmrs}{DMRS}{DeModulation Reference Signal}
\newacronym{e2e}{E2E}{End-to-End}
\newacronym{ppp}{PPP}{Poission Point Process}
\newacronym{si}{SI}{Study Item}
\newacronym{ecn}{ECN}{Explicit Congestion Notification}
\newacronym{edf}{EDF}{Earliest Deadline First}
\newacronym{enb}{eNB}{eNodeB}
\newacronym{epc}{EPC}{Evolved Packet Core}
\newacronym{es}{ES}{Edge Server}
\newacronym{cav}{CAV}{Connected and Autonomous Vehicle}
\newacronym{fdma}{FDMA}{Frequency Division Multiple Access}
\newacronym{fdd}{FDD}{Frequency Division Duplexing}
\newacronym{upa}{UPA}{Uniform Planar Array}
\newacronym[firstplural=Radio Access Technologies (RATs)]{rat}{RAT}{Radio Access Technology}
\newacronym[firstplural=Radio Access Technology (RTs)]{rt}{RT}{Radio Technology}
\newacronym{fs}{FS}{Fast Switching}
\newacronym{isd}{ISD}{inter-site distance}
\newacronym{ftp}{FTP}{File Transfer Protocol}
\newacronym{gnb}{gNB}{Next Generation Node Base}
\newacronym{harq}{HARQ}{Hybrid Automatic Repeat reQuest}
\newacronym{hetnet}{HetNet}{Heterogeneous Network}
\newacronym{hh}{HH}{Hard Handover}
\newacronym{hol}{HOL}{Head-of-Line}
\newacronym{ia}{IA}{Initial Access}
\newacronym{imt}{IMT}{International Mobile Telecommunication}
\newacronym{iot}{IoT}{Internet of Things}
\newacronym{los}{LOS}{Line of Sight}
\newacronym{lte}{LTE}{Long Term Evolution}
\newacronym{m2m}{M2M}{Machine to Machine}
\newacronym{mac}{MAC}{Medium Access Control}
\newacronym{mc}{MC}{Multi-Connectivity}
\newacronym{mcs}{MCS}{Modulation and Coding Scheme}
\newacronym{mec}{MEC}{Mobile Edge Cloud}
\newacronym{mi}{MI}{Mutual Information}
\newacronym{mimo}{MIMO}{Multiple Input Multiple Output}
\newacronym{mmwave}{mmWave}{millimeter wave}
\newacronym{mptcp}{MPTCP}{Multipath TCP}
\newacronym{mr}{MR}{Maximum Rate}
\newacronym{mss}{MSS}{Maximum Segment Size}
\newacronym{mse}{MSE}{Mean Square Error}
\newacronym{mtd}{MTD}{Machine-Type Device}
\newacronym{mtu}{MTU}{Maximum Transmission Unit}
\newacronym{nfv}{NFV}{Network Function Virtualization}
\newacronym{nlos}{NLOS}{Non Line of Sight}
\newacronym{nlosb}{NLOSb}{Building Non Line of Sight}
\newacronym{nlosv}{NLOSv}{Vehicle Non Line of Sight}
\newacronym{nr}{NR}{New Radio}
\newacronym{ofdm}{OFDM}{Orthogonal Frequency Division Multiplexing}
\newacronym{pdcch}{PDCCH}{Physical Downlonk Control Channel}
\newacronym{pdcp}{PDCP}{Packet Data Convergence Protocol}
\newacronym{pdsch}{PDSCH}{Physical Downlink Shared Channel}
\newacronym{pdu}{PDU}{Packet Data Unit}
\newacronym{pf}{PF}{Proportional Fair}
\newacronym{pgw}{PGW}{Packet Gateway}
\newacronym{phy}{PHY}{Physical}
\newacronym{pbch}{PBCH}{Physical Broadcast Channel}
\newacronym[plural=\gls{mme}s,firstplural=Mobility Management Entities (MMEs)]{mme}{MME}{Mobility Management Entity}
\newacronym{prb}{PRB}{Physical Resource Block}
\newacronym{pss}{PSS}{Primary Synchronization Signal}
\newacronym{pucch}{PUCCH}{Physical Uplink Control Channel}
\newacronym{pusch}{PUSCH}{Physical Uplink Shared Channel}
\newacronym{rach}{RACH}{Random Access Channel}
\newacronym{ran}{RAN}{Radio Access Network}
\newacronym{red}{RED}{Random Early Detection}
\newacronym{rf}{RF}{Radio Frequency}
\newacronym{rlc}{RLC}{Radio Link Control}
\newacronym{rlf}{RLF}{Radio Link Failure}
\newacronym{rrc}{RRC}{Radio Resource Control}
\newacronym{rrm}{RRM}{Radio Resource Management}
\newacronym{rr}{RR}{Round Robin}
\newacronym{rs}{RS}{Remote Server}
\newacronym{rsrp}{RSRP}{Reference Signal Received Power}
\newacronym{rss}{RSS}{Received Signal Strength}
\newacronym{rtt}{RTT}{Round Trip Time}
\newacronym{rw}{RW}{Receive Window}
\newacronym{rx}{RX}{Receiver}
\newacronym{sa}{SA}{standalone}
\newacronym{sack}{SACK}{Selective Acknowledgment}
\newacronym{sap}{SAP}{Service Access Point}
\newacronym{sch}{SCH}{Secondary Cell Handover}
\newacronym{scoot}{SCOOT}{Split Cycle Offset Optimization Technique}
\newacronym{sdma}{SDMA}{Spatial Division Multiple Access}
\newacronym{sinr}{SINR}{Signal to Interference plus Noise Ratio}
\newacronym{sm}{SM}{Saturation Mode}
\newacronym{snr}{SNR}{Signal to Noise Ratio}
\newacronym{psnr}{PSNR}{Peak Signal to Noise Ratio}
\newacronym{son}{SON}{Self-Organizing Network}
\newacronym{ss}{SS}{Synchronization Signal}
\newacronym{srs}{SRS}{Sounding Reference Signal}
\newacronym{sss}{SSS}{Secondary Synchronization Signal}
\newacronym{tb}{TB}{Transport Block}
\newacronym{tcp}{TCP}{Transmission Control Protocol}
\newacronym{tdd}{TDD}{Time Division Duplexing}
\newacronym{tdma}{TDMA}{Time Division Multiple Access}
\newacronym{tfl}{TfL}{Transport for London}
\newacronym{tm}{TM}{Transparent Mode}
\newacronym{prr}{PRR}{Packet Reception Ratio}
\newacronym{trp}{TRP}{Transmitter Receiver Pair}
\newacronym{tti}{TTI}{Transmission Time Interval}
\newacronym{ttt}{TTT}{Time-to-Trigger}
\newacronym{tx}{TX}{Transmitter}
\newacronym{ue}{UE}{User Equipment}
\newacronym{ul}{UL}{Uplink}
\newacronym{uml}{UML}{Unified Modeling Language}
\newacronym{um}{UM}{Unacknowledged Mode}
\newacronym{utc}{UTC}{Urban Traffic Control}
\newacronym{vm}{VM}{Virtual Machine}
\newacronym{rsrq}{RSRQ}{Reference Signal Received Quality}
\newacronym{rssi}{RSSI}{Received Signal Strength Indicator}
\newacronym{crs}{CRS}{Cell Reference Signal}
\newacronym{v2v}{V2V}{Vehicle-to-Vehicle}
\newacronym{v2i}{V2I}{Vehicle-to-Infrastructure}
\newacronym{v2n}{V2N}{Vehicle-to-Network}
\newacronym{v2x}{V2X}{Vehicle-to-Everything}
\newacronym{vn}{VN}{Vehicular Node}
\newacronym{dsrc}{DSRC}{Dedicated Short Range Communication}
\newacronym{ci}{CI}{context information}
\newacronym{voi}{VoI}{value of information}
\newacronym{gps}{GPS}{Global Positioning System}
\newacronym{qos}{QoS}{Quality of Service}
\newacronym{qoe}{QoE}{Quality of Experience}
\newacronym{ml}{ML}{Machine Learning}
\newacronym{ahp}{AHP}{Analytic Hierarchy Process}
\newacronym{lidar}{LiDAR}{Light Detection and Ranging}
\newacronym{sumo}{SUMO}{Simulation of Urban MObility}
\newacronym{wave}{WAVE}{Wireless Access in Vehicular Environment}
\newacronym{c-its}{C-ITS}{Connected Intelligent Transportation System}
\newacronym{dash}{DASH}{Dynamic Adaptive Streaming over HTTP}
\newacronym{http}{HTTP}{HyperText Transfer Protocol}
\newacronym{3d}{3D}{thee-dimensional}
\newacronym{dnn}{DNN}{Deep Neural Network}
\newacronym{gpcc}{G-PCC}{geometry-based point cloud compression}
\newacronym{hsc}{HSC}{Hybrid Semantic Compression}
\newacronym{bpp}{BPP}{bits per point}
\newacronym{cpu}{CPU}{Central Processing Unit}
\newacronym{iou}{IoU}{Intersection over Union}

\graphicspath{{./figures/}}
\newcommand{\mytexttilde}{{\raise.17ex\hbox{$\scriptstyle\mathtt{\sim}$}}}

\IEEEoverridecommandlockouts
\newcommand\copyrightnotice{%
\begin{tikzpicture}[remember picture,overlay]
\node[anchor=south,yshift=10pt] at (current page.south) {\fbox{\parbox{\dimexpr\textwidth-\fboxsep-\fboxrule\relax}{
\footnotesize \textcopyright 2021 IEEE. Personal use of this material is permitted.
Permission from IEEE must be obtained for all other uses, in any current or future media,
including reprinting/republishing this material for advertising or promotional purposes,
creating new collective works, for resale or redistribution to servers or lists,
or reuse of any copyrighted component of this work in other works.}}};
\end{tikzpicture}
}

\title{Hybrid Point Cloud Semantic Compression for Automotive Sensors: A Performance Evaluation}

\author{Andrea Varischio, Francesco Mandruzzato, Marcello Bullo \\ Marco Giordani, Paolo Testolina, Michele Zorzi \\ Department of Information Engineering, University of Padova, Padova, Italy \\ \ \texttt{\{name.surname\}@dei.unipd.it} }

\begin{document}

\maketitle
\copyrightnotice

\begin{abstract}
In a fully autonomous driving framework, where vehicles operate without human intervention, information sharing plays a fundamental role.
In this context, new network solutions have to be designed to handle the large volumes of data generated by the rich sensor suite of the cars in a reliable and efficient way. 
Among all the possible sensors,  Light Detection and Ranging (LiDAR) can produce an accurate 3D point cloud representation of the surrounding environment, which in turn generates high data rates.
For this reason, efficient point cloud compression is paramount to alleviate the burden of data transmission over bandwidth-constrained channels and to facilitate real-time communications.
In this paper, we propose a pipeline to efficiently compress LiDAR observations in an automotive scenario.
First, we leverage the capabilities of RangeNet++, a Deep Neural Network (DNN) used to semantically infer point labels, to  reduce the channel load by selecting the most valuable environmental data to be disseminated.
Second,  we compress the selected points  using \mbox{Draco}, a 3D compression algorithm which is able to obtain compression up to the quantization error.
Our experiments, validated on the Semantic KITTI dataset, demonstrate that it is possible to compress and send the information at the frame rate of the LiDAR, thus achieving \mbox{real-time} performance.
\end{abstract}

\IEEEkeywords
Autonomous driving, compression, Draco, Deep Neural Networks (DNNs), RangeNet++, point cloud,  LiDAR.
\endIEEEkeywords

\begin{tikzpicture}[remember picture,overlay]
\node[anchor=north,yshift=-10pt] at (current page.north) {\parbox{\dimexpr\textwidth-\fboxsep-\fboxrule\relax}{
\centering\footnotesize This paper has been accepted for presentation at IEEE International Conference on Communications (ICC) 2021. \textcopyright 2021 IEEE.\\
Please cite it as: A. Varischio, F. Mandruzzato, M. Bullo, M. Giordani, P. Testolina, M. Zorzi, "Hybrid Point Cloud Semantic Compression for Automotive Sensors: A Performance Evaluation," IEEE International Conference on Communications (ICC), Virtual/Montreal, Canada, 2021}};
\end{tikzpicture}

\section{Introduction}
\label{sec:introduction}
\gls{lidar} sensors play a crucial role in the field of autonomous driving, providing vehicles with a \gls{3d} omnidirectional perception of the surrounding environment~\cite{lu2014connected}.
Compared to other types of sensors, such as automotive cameras and radars, LiDAR data is represented in the form of large point clouds carrying both geometry and attribute information (e.g., depth properties), to achieve fine detection, localization, and recognition of the surrounding objects~\cite{chen2017multi}.
However,  LiDAR point clouds may generate very large volumes of data. For example, a 3D frame with 0.7 million points acquired at 30 fps needs a bandwidth of around 500 Mbps~\cite{cao20193d}, which can be challenging to handle with standard communication systems~\cite{7786130} (e.g., the IEEE 802.11p transmission service, currently employed for \gls{v2v} operations, offers data exchange at a nominal rate up to 27 Mbps~\cite{giordani2018feasibility}).

One possible approach to solve capacity issues is by developing new radio systems, e.g., operating in the lower part of the \gls{mmwave} spectrum, as currently promoted by  IEEE and 3GPP standardization activities for next-generation vehicular networks~\cite{zugno2020toward}.
This potential is however hindered by the harsh propagation characteristics of the above-6 GHz bands, which might be particularly critical in the dynamic vehicular environment~\cite{MOCAST_2017}.

Data rates can be further reduced by compressing the point clouds in a sufficient ratio and make them suitable for real-time transmission.
In this context, 2D-oriented compression techniques, such as JPG or PNG, might be ineffective in the 3D space as point clouds, unlike images, are sparse and non-grid structured.
The most common way to structure 3D data is indeed to enclose points in a bounding cube and discretize the content in a so-called \emph{voxel-grid}, thus obtaining a three-dimensional matrix at a specific resolution.

\begin{figure*}
 \center
\setlength{\belowcaptionskip}{-0.18cm} 
  \includegraphics[width=\textwidth]{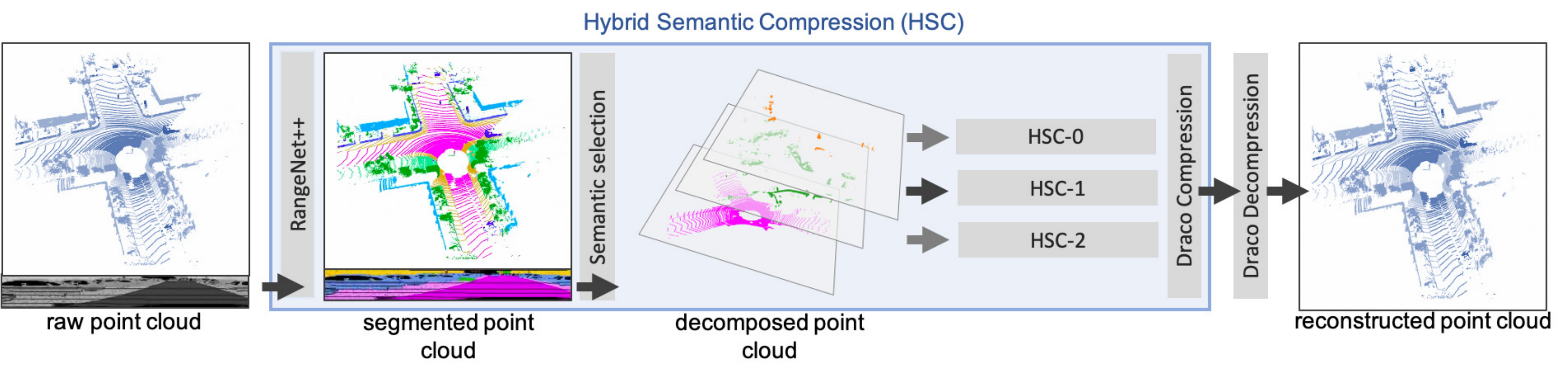}
  \caption{An overview of the proposed Hybrid Semantic Compression (HSC)  pipeline for LiDAR point clouds.}
  \label{fig:pipeline}
\end{figure*}

Following this intuition, several literature works have approached 3D compression by using data structures such as Octrees~\cite{schnabel2006octree}, which could not however detect and appropriately remove redundant information hidden in the point cloud representations.
More recently, novel techniques based on  deep learning  have made it possible to further improve point cloud compressibility by extracting features from individual points and predicting voxel occurrence in 3D scenes, e.g., ~\cite{huang2020octsqueeze,qi2017pointnet}.

Based on the above introduction, in this paper we aim at improving the compression efficiency of LiDAR point clouds by combining the compression capabilities of the Google software~Draco~\cite{Draco} with the semantic segmentation functionalities provided by RangeNet++~\cite{rnet}, a \gls{dnn} used to assign class labels to data points.
As illustrated in Fig.~\ref{fig:pipeline}, the proposed \gls{hsc} pipeline works as follows:
\begin{enumerate}
  \item We first infer full semantic segmentation of LiDAR point clouds with RangeNet++, so as to identify the most valuable objects in the scene (typically pedestrians and vehicles). If capacity is limited, this allows deferring or canceling the transmissions of the least critical acquisitions~\cite{giordani2019investigating,higuchi2019value}, while prioritizing the most important LiDAR frames.
  \item We compress the resulting point cloud with Draco, which defines 15 quantization levels to consider different encoding/decoding speeds and discretization losses.
\end{enumerate}

The performance of the proposed \gls{hsc} framework is evaluated in terms of average compression ratio (an indication of how efficiently data is compressed), \gls{psnr}, and symmetric Chamfer Distance (used to measure the quality of the reconstructed point cloud). For completeness, our approach is compared with a baseline \gls{gpcc} scheme which is able to handle sparse data working directly on the 3D space~\cite{graziosi2020overview}. 
Our results, which are validated on the Semantic KITTI dataset~\cite{behley2019semantickitti}, show that HSC can deliver quasi-real-time compression performance with limited accuracy degradation, especially when designing aggressive  semantic segmentation.

The rest of this paper is organized as follows. Sec.~\ref{sec:related_work} concisely describes related works on point cloud compression.
Sec.~\ref{sec:model} and Sec.~\ref{sec:Draco} provide a brief overview of RangeNet++ and Draco, respectively.
In Sec.~\ref{sec:results} we present our  experimental results, whereas Sec.~\ref{sec:conclusions} concludes the article.

\section{Related Work}
\label{sec:related_work}

The adoption of LiDARs on commercial vehicles triggered new interest in 3D point cloud compression.
Prior works, e.g., \cite{wu2018squeezeseg,xu2020squeezesegv3},  first addressed this research challenge by projecting the 3D point cloud into a 2D representation to perform  bi-dimensional compression, achieving \mbox{real-time} performance.
Despite their desirable computational speed, however, such methods were proven to be inefficient when applied to dense 360-degree LiDAR~point clouds.

In recent times, the introduction of deep learning in networks allowed researchers to broaden their horizons, and design more sophisticated compression schemes based on the  automatic extraction of features from data.
In \cite{zhou2018voxelnet}, for example, Zhou \emph{et al.} introduced an end-to-end trainable deep network to extract features automatically from equally spaced 3D voxel structures, and to make bounding box prediction based on them. 
However, the sparsity of the 3D LiDAR data resulted in many empty voxel cells, thus making this representation inefficient.
In \cite{qi2017pointnet} the authors designed a neural network to process point clouds while being invariant to input permutations.
Despite its inherent simplicity, this method performs poorly on large LiDAR point clouds, and the memory required to store such representations is prohibitively large.
Some other studies have introduced auto-encoders or recurrent neural networks that directly work on the 3D geometry. Specifically, in \cite{balle2018variational} variational auto-encoders, in combination with a scale hyperprior, have been exploited to capture spatial dependencies in the latent space. The drawback of such models is that they require very complex learning capabilities, especially with sparse data structures, which may be difficult to guarantee on board of~vehicles.

Research efforts have then been dedicated to the design of compression strategies operating on data structures able to accurately represent point clouds.
One of the most popular methods is based on tree structures, like \mbox{Octree} or \mbox{KD-tree}.
The advantage is their ability to model arbitrary point clouds in a hierarchical way, dealing also with sparsity.
For example, in~\cite{huang2020octsqueeze} point clouds are first stored in an \mbox{Octree} and then the extracted context features are fed into a deep entropy model to be encoded into a compact bitstream.

Compression can be further improved when a full characterization of the environment is provided, as described in~\cite{rnet,zhang2020polarnet,cortinhal2020salsanext}, which is the main idea behind the proposed HSC framework.


\section{RangeNet++}
\label{sec:model}

In this section  we provide a brief description of RangeNet++~\cite{rnet}, which is used to assign a class label to 3D LiDAR points, an essential step in HSC to filter the raw point cloud keeping only the most relevant/critical points.
For a more complete description of the RangeNet++ network  we refer the interested reader to~\cite{rnet}.

\mbox{RangeNet++} is a deep convolutional neural network able to accurately perform semantic segmentation using only the LiDAR point cloud and its reflectance. 
Furthermore, it can infer the point labels without discarding any point, while working at the frame rate of the LiDAR scanner.

The raw 3D data is first mapped on a range image through a spherical projection.
Features are then extracted from the range representation and fed into an encoder-decoder architecture.
After inferring the labels (e.g., road, pedestrian, vegetation, etc.) for the current frame, each~point is reprojected to the original point cloud.
A \mbox{k-nearest} neighbor (kNN) search is then employed to reduce the blurring introduced by the network and increase the segmentation~accuracy.

As reported in \cite{rnet}, the size of the range image affects the accuracy of the segmentation and the number of frames processed  per second. 
In this work, we set the size of the image resolution to $[64\times2048]$, which is the maximum resolution supported by RangeNet++, and was reported to provide very accurate results even with limited power resources~\cite{rnet}.

\section{Draco}
\label{sec:Draco}

LiDAR acquisitions are very sparse point clouds without a pre-defined structure.
In this context, Draco~\cite{Draco}, a software designed by Google, uses a KD-tree data structure~\cite{geoemtry_draco} to efficiently organize 3D data while dealing with sparsity,  and provides compression at very high speed.

Despite its wide adoption in the literature, the software lacks a detailed documentation. In this work, we just give an idea of how it works.
Unlike Octree-based solutions, Draco keeps on splitting the point cloud from the center, alternating the axes at each iteration  in order to build a more balanced space-partitioning tree.
The number of points after each split operation is encoded as the difference between the number of points contained in the smaller half of the point cloud and the total number of points, divided by two.
This results in more leading zeros in the encoded values, which are better compressed by arithmetic encodings.

Draco's flexibility allows to define 15 quantization levels $(0-14)$ and 11 compression levels $(0-10)$ that trade off the compression efficiency against its speed.
Compression accuracy is then proportional to the quantization error: as the quantization level grows, the compression quality increases at the cost of a larger file, whereas higher quantization levels translate in more compression but lower decompression speed.

\section{Compression Performance Results}
\label{sec:results}

\label{sec:results}

In this section we first overview our performance metrics (Sec.~\ref{ssec:metrics}), then  describe the G-PCC architecture, which is used to benchmark the performance of the proposed HSC framework (Sec.~\ref{ssec:gpcc}), and finally  present our main performance results (Sec.~\ref{ssec:results}).

\begin{figure}[t!] 
  \center{\includegraphics[width=0.8\linewidth]{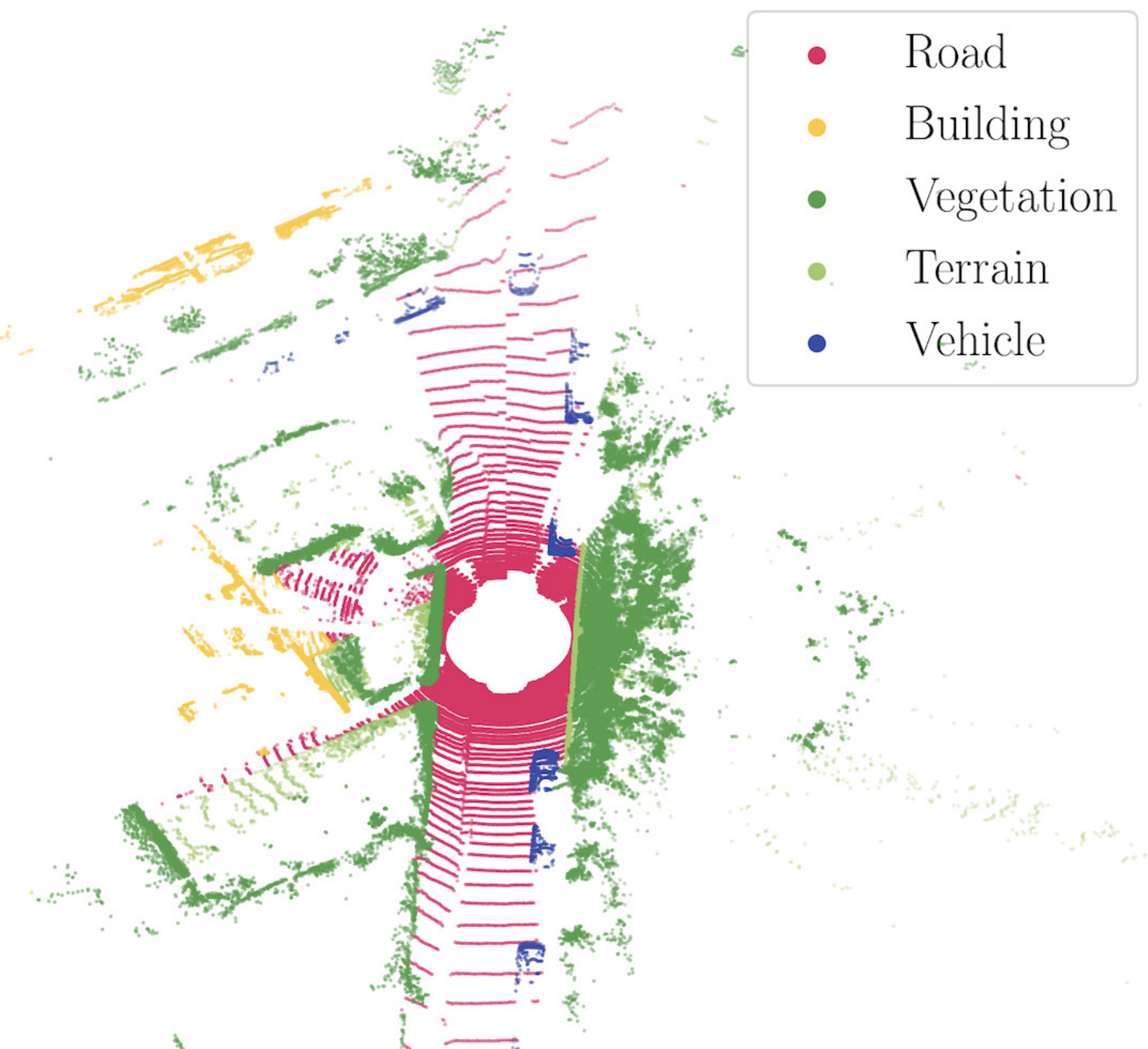}}
  \caption{Example of point cloud semantic segmentation performed by \mbox{RangeNet++}. Each set of points may or may not be removed from the raw LiDAR point  cloud before compression, depending on how aggressively the HSC framework is configured.}
  \label{fig:test}
\end{figure}

To evaluate the compression performance of  HSC  we use the Semantic KITTI dataset \cite{behley2019semantickitti}, where LiDAR sensors generate point clouds of 120'000 points per frame, on average every 100 ms.
HSC has been designed to support three possible transmission levels:
\begin{itemize}
  \item \textbf{HSC-0}: The raw LiDAR acquisition is immediately compressed by Draco and sent through the channel. In this case, RangeNet++ is not used.
  \item \textbf{HSC-1}:  RangeNet++ is used to perform semantic segmentation on the raw LiDAR acquisitions, as depicted in Fig.~\ref{fig:test}. After that, the points belonging to the road elements are removed from the point cloud, thus reducing the file size. This choice can be justified by assuming the availability of topology information from auxiliary data sources such as Google Maps or similar mapping tools.  
  \item \textbf{HSC-2}: Compared to HSC-1, buildings, vegetation, and traffic signs, which may  be provided by camera~sensors, if available, are also removed from the point cloud after RangeNet++ segmentation. Consequently, the resulting point cloud to be compressed consists only of dynamic elements like pedestrians and vehicles, i.e., the most critical elements in autonomous driving scenarios~\cite{giordani2019framework}.
\end{itemize}


It is important to highlight the versatility of the proposed pipeline: the segmentation performed by \mbox{RangeNet++} can be exploited during the transmission phase to prioritize safety-critical over less relevant information.
Consider, for instance, the \gls{udp}, a common choice at the transport layer for time-sensitive applications.
UDP's lean design permits fast data packet transmissions, while in turn suffering  from lossy connection, which may require full retransmission of packets.
Moreover, UDP has a maximum packet size of 65'535 Bytes, which poses a significant constraint on the dimension of the LiDAR data to avoid fragmentation, even after compression. 
In light of this, data selection plays a vital role in this context.
In the HSC framework, the opportunity to remove from raw LiDAR point clouds nonessential static elements after RangeNet++ segmentation (i.e., in HSC-1 and HSC-2) makes it possible to encode  higher-priority information in the same UDP packets, which are to be transmitted first.

\subsection{Performance Metrics} 
\label{ssec:metrics}

In this section we present the metrics used to evaluate the  performance of the investigated compression solutions.
\smallskip

\paragraph{\Gls{bpp}}
This quantity represents  the number of bits used to compress each point in the original point cloud, and is computed as the ratio between the compressed file size and the number of points  in the compressed file. 
A larger BPP  would result in a better quality of the reconstructed point cloud, at the cost of a lager memory~footprint.

\paragraph{{Average compression ratio}}
This metric indicates the relative reduction in size of the data representation produced by  compression, and is measured as the ratio between the raw point cloud  size and the  compressed file size.

\paragraph{{Symmetric point-to-point Chamfer Distance}}
To evaluate the quality of the reconstructed point cloud $\hat{P}$ with respect to the raw point cloud $P$, we adopt the symmetric point-to-point Chamfer Distance $\mathrm{CD}_{\rm sym}$, which is defined as 
\medmuskip=0mu
\thickmuskip=0mu
\begin{equation}
\begin{aligned}
\mathrm{CD_{\rm sym}} = \sum_{\forall \mathbf{p}\in P} \min_{\forall \hat{\mathbf{p}}\in\hat{P}} \lVert \mathbf{\mathbf{p}} - \hat{\mathbf{p}} \rVert_2^2 + \sum_{\forall \hat{\mathbf{p}}\in \hat{P}} \min_{\forall \mathbf{p}\in P} \lVert \mathbf{ \mathbf{p}} - \hat{\mathbf{p}} \rVert_2^2.
\end{aligned}
\end{equation}
\medmuskip=6mu
\thickmuskip=6mu
As $\mathrm{CD}_{\rm sym}$ approaches zero, the reconstructed point cloud, $\hat{P}$, better approximates the original one, $P$. 
\smallskip

\paragraph{{Point-to-plane \gls{psnr}}}
Given a point $\bm{p}\in P$ and its nearest neighbor $\bm{q}\in \hat{P}$, the point-to-plane \gls{mse} computed with respect to $P$ is
\begin{equation}
	\mathrm{MSE}_{P\to \hat{P}} = \dfrac{1}{|P|}\sum_{\forall \bm{p}\in P} \left( \left\langle \bm{p}-\bm{q}, \bm{n}_{\bm{q}} \right\rangle \right)^2,
\label{eq:mse}
\end{equation}
where $\bm{n}_{\bm{q}}$ is the surface normal in $\bm{q}\in \hat{P}$, $|P|$ is the cardinality of the point cloud $P$ and $\left\langle \bm{p}-\bm{q}, \bm{n}_{\bm{q}} \right\rangle$ is the projection of vector $\bm{p}-\bm{q}$ on $\bm{n}_{\bm{q}}$. Then, the point-to-plane \glsentryfull{psnr} with respect to $P$ is given by
\begin{equation}
	\mathrm{PSNR}_{P\to \hat{P}} = 10 \log_{10}\left(\frac{(\gamma_P^*)^2}{\mathrm{MSE}_{P\to \hat{P}}}\right)
\label{eq:psnr-ref}
\end{equation}
where $\gamma_P^*$ is a peak value in the point cloud $P$.
Conventionally, the peak value $\gamma_P^*$ is defined based on the diagonal distance of a bounding box of the point cloud $P$, which would however imply that higher \gls{psnr} is produced when larger objects are considered, thus resulting in an unbalanced metric.
To solve this issue, we compute $\gamma_P^*$ using the intrinsic resolution of the point cloud $P$, as proposed in~\cite{psnr}.
In practice, $\gamma_P^*$ is derived from the nearest neighbor distances $d_p$ for all points $\mathbf{p}$ in the input point cloud~$P$, i.e., $\gamma_P^* = \max_{\forall\bm{p} \in P}\left\{ d_p \right\}$.
Finally, the symmetric \gls{psnr} between $P$ and $\hat{P}$ is defined as
\begin{equation}
	\mathrm{PSNR}_{P,\hat{P}} = \min\left\{{\mathrm{PSNR}_{P\to \hat{P}},\ \mathrm{PSNR}_{\hat{P}\to P}}\right\}.
\label{eq:psnr}
\end{equation}


\subsection{Compression Baseline: G-PCC}
\label{ssec:gpcc}

The MPEG group is developing a new standard for point cloud compression called \gls{gpcc} (also known as TMC13)~\cite{graziosi2020overview}. The G-PCC codec algorithm, like Draco, works directly in the 3D space and allows to code both geometry and attributes of a point cloud following two parallel approaches. 
In particular, G-PCC voxelizes the raw point cloud by quantizing coordinates to closest integer. 
Furthermore, translation and scaling transformations are applied such that the resulting point cloud resides in a bounding cube of size $[0,2^d]^3$, for some non negative integer $d$ defined as:
\medmuskip=2mu
\thickmuskip=2mu
\begin{equation}
d = \left\lceil(\log_2 \left[\max(x_n^{{\rm int}},y_n^{\rm int},z_n^{\rm int})+1)\right]\right\rceil, \: \forall n=1,\dots,N,
\end{equation}
\medmuskip=6mu
\thickmuskip=6mu
where $[x_n^{{\rm int}},y_n^{\rm int},z_n^{\rm int}]$ are the point cloud's 3D coordinates, $N$ is the total number of points in the cloud, and $\left\lceil \cdot \right\rceil$ is the ceiling function.
The processed point clouds are then stored in an Octree-based data structure and finally arithmetic coding is applied.
Notice that the quantization performed by G-PCC may collapse many points into the same position depending on the spatial resolution allowed in the voxel grid.
On this work, the scaling factor has been tuned so duplicate points are avoided.

\subsection{Experimental Results}
\label{ssec:results}

In the following results we compare the compression efficiency (in terms of file size, compression ratio and  processing time) and accuracy (in terms of Chamfer Distance and PSNR) of HSC vs. a baseline G-PCC implementation.
All our tests were performed on an NVIDIA\textsuperscript{TM} TITAN\textsuperscript{\textregistered} RTX GPU, equipped with an Intel\textsuperscript{\textregistered} Core\textsuperscript{TM} i7-7700K CPU at 4.20 GHz and 32 GB of RAM.
The metrics are shown as a function of the BPP used to compress each point in the raw point cloud, as defined in Sec. \ref{ssec:metrics}.
Specifically, the value of the BPP depends on how aggressive the compression is, i.e., on the adopted quantization level (for HSC) and the voxelization scaling factor (for G-PCC).
\smallskip

\begin{figure}[t!] 
    \centering 
    \includegraphics[width=0.88\columnwidth]{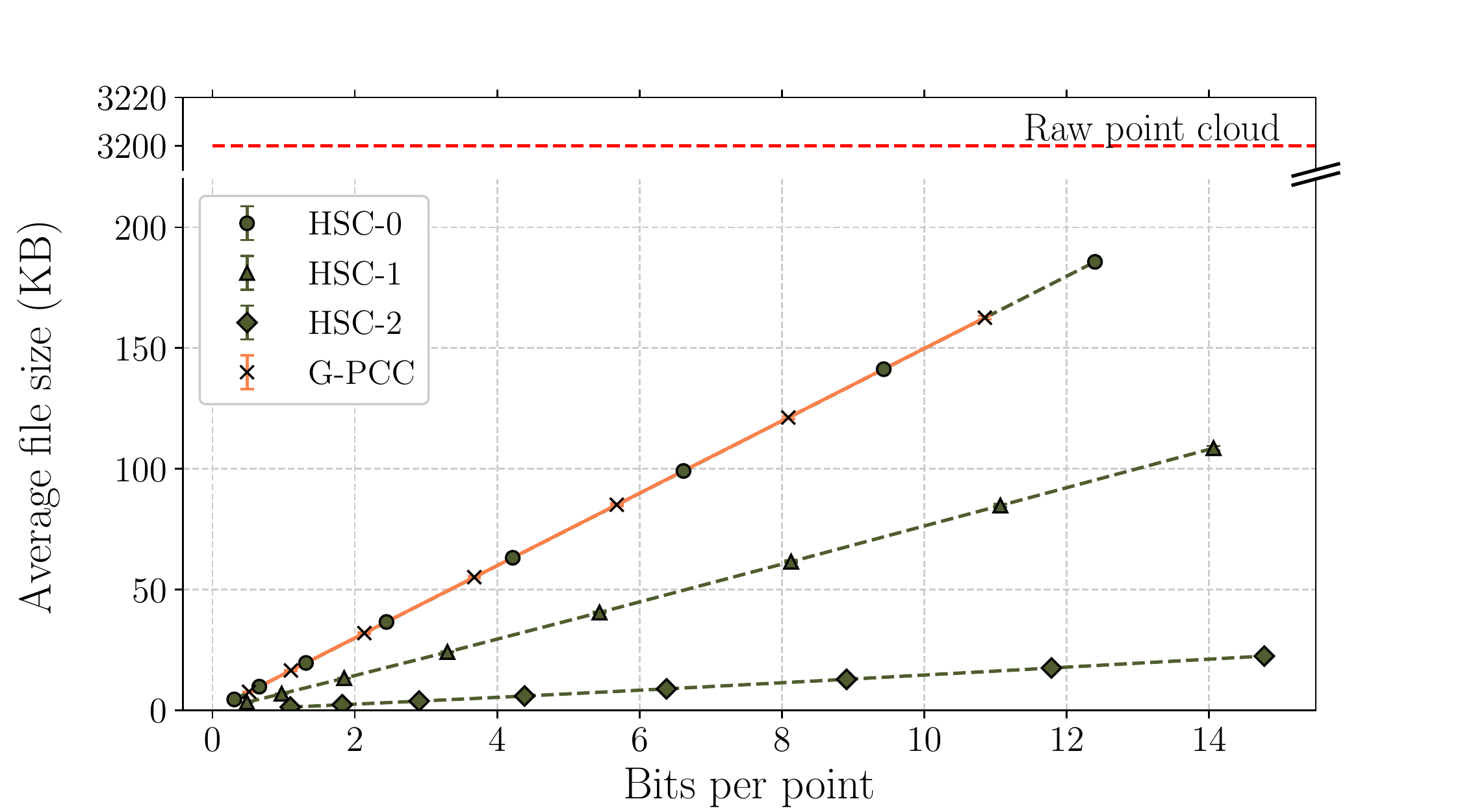}
    \caption{File size vs. the BPP, as a function of the compression strategy.}%
    \label{fig:filesize}
\end{figure}

\begin{figure}[t!]
    \centering
    \includegraphics[width=0.88\columnwidth]{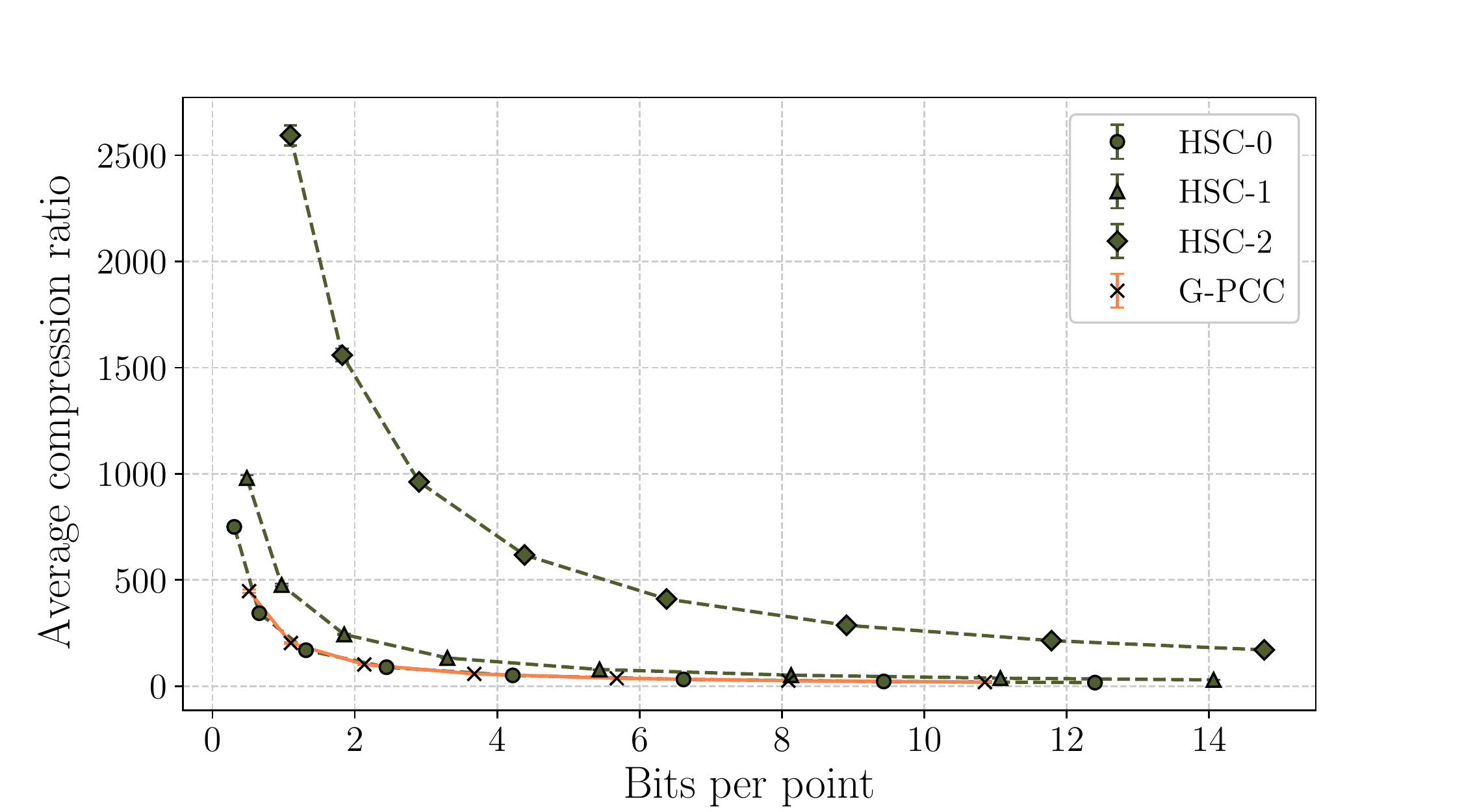}
    \caption{Average compression ratio vs. the BPP, as a function of the compression strategy.}%
    \label{fig:compr_ratio}
\end{figure}

\textbf{Compression efficiency.}
In Fig. \ref{fig:filesize} we show the average file size obtained with HSC and G-PCC. 
In general, it is necessary to reduce the file size as much as possible to keep the channel free for transmissions generated from sensory sources other than the LiDAR, or to serve high-priority traffic requests (e.g., in case of emergency). 
It is evident that, in fact,  even considering only LiDAR transmissions, standard vehicular communication technologies  would not be able to satisfy traffic requests efficiently, thus creating congestion: from Fig.~\ref{fig:filesize}, raw LiDAR perceptions of 3200 KB, generated every 100 ms, would require a data rate of at least 256 Mbps,  which is far beyond what current IEEE 802.11p implementations for V2V communications can support.
Notably, the HSC framework can compress the raw LiDAR point cloud  up to 8 times more effectively than G-PCC. 
Moreover, unlike Draco (i.e., HSC-0) and G-PCC, HSC-2 allows to encapsulate the semantically inferred point cloud in a single UDP packet of 65'535 Bytes regardless of the quantization level, a fundamental pre-requisite to guarantee that the most critical and time-sensitive data reaches the receiver(s) with the lowest latency and the highest reliability.
Similarly, this is possible using HSC-1 with less than 10 BPP.

The same conclusions can be derived from Fig.~\ref{fig:compr_ratio}, which illustrates the average compression ratio vs. the BPP. As expected, not only can HCP provide better compression compared to \mbox{G-PCC}, but semantic segmentation before compression (i.e., HSC-1 and HSC-2) can also improve the compression efficiency by more than 10 times with respect to a standard Draco-only approach (i.e., HSC-0) by removing non-critical data before transmission.

\begin{figure}[t!]
\centering
        \begin{subfigure}[b]{0.49\textwidth}   
            \centering 
            \includegraphics[width=0.9\columnwidth]{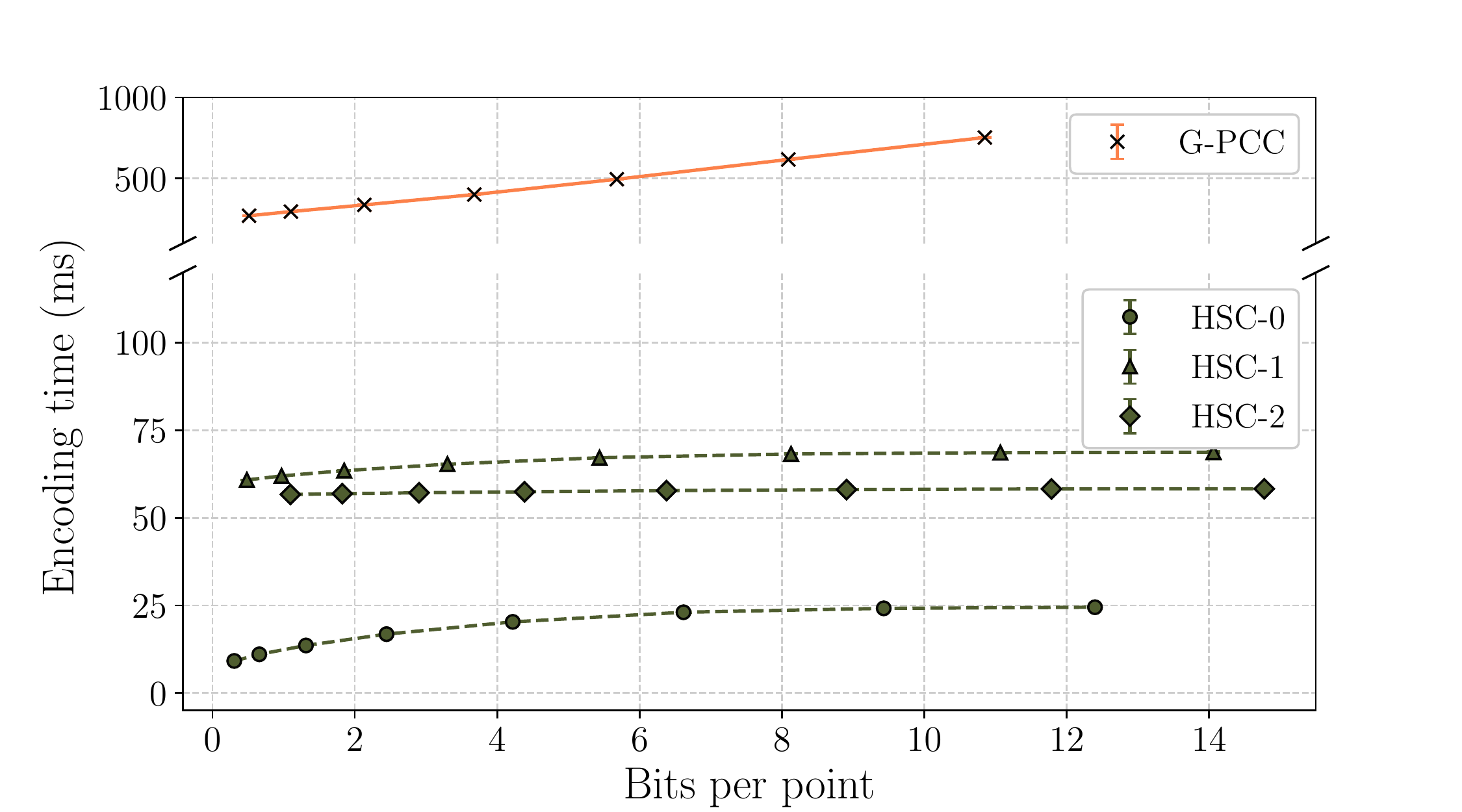}
            \caption{Encoding time.}%
            \label{fig:encoding}
        \end{subfigure}\\\vspace{0.33cm}
        \begin{subfigure}[b]{0.49\textwidth}   
            \centering 
            \includegraphics[width=0.9\columnwidth]{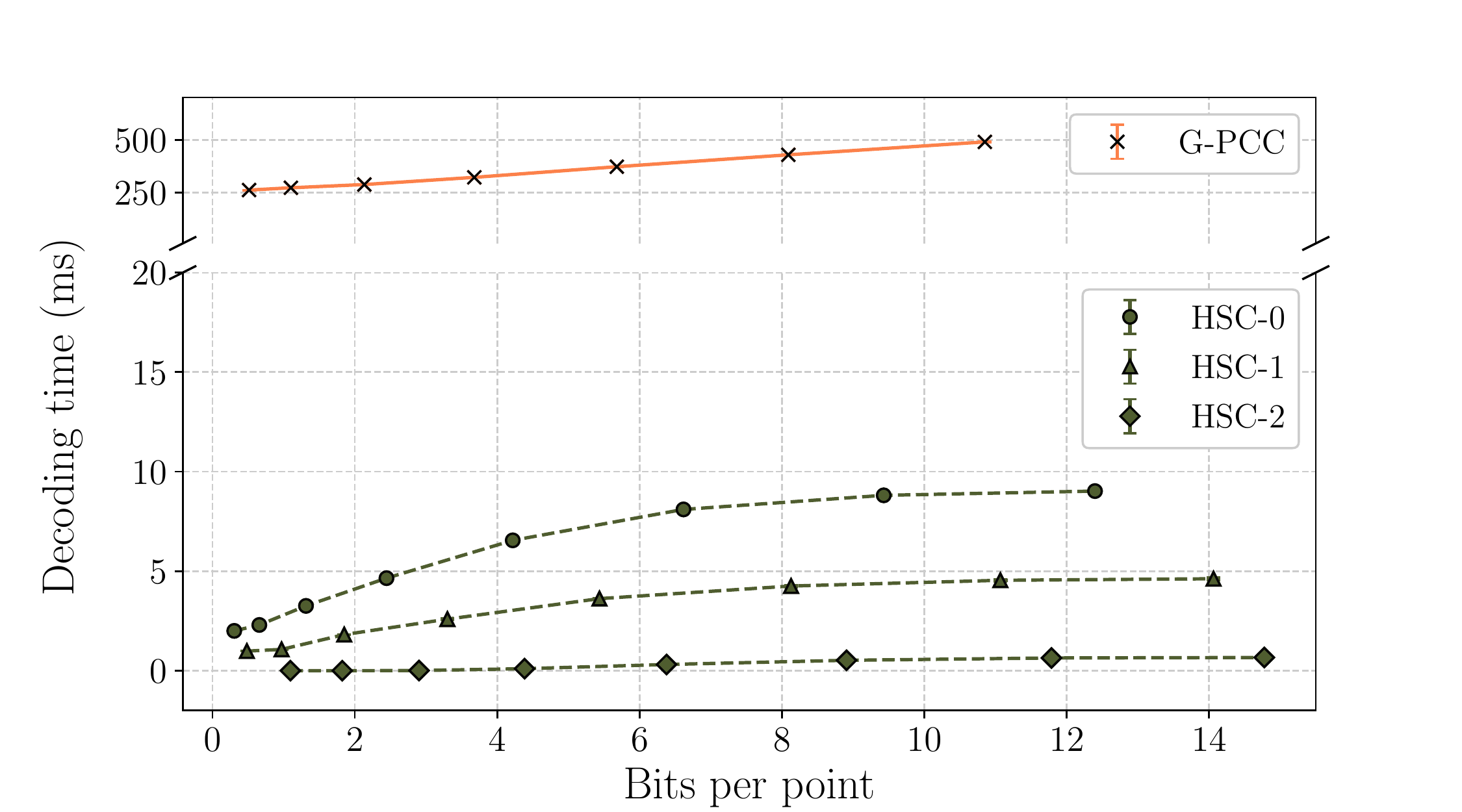}
            \caption{Decoding time. }%
            \label{fig:decoding}
        \end{subfigure}
        \caption{Encoding and decoding time of the investigated compression strategies vs. the BPP. For HSC-1 and HSC-2, these metrics also account for the average time required by RangeNet++ to semantically infer labels to point clouds.}
        \label{fig:CPU_time}
    \end{figure}

In terms of computational efficiency, Fig.~\ref{fig:CPU_time} shows that, as the BPP increases, the quantization is proportionally less aggressive and, consequently, both encoding and decoding times increase significantly.
Regarding \mbox{G-PCC}, the time required to encode and decode a point cloud is orders of magnitude higher than in HSC, which validates the inefficiency of representing 3D data in an Octree-based  structure compared to a KD-tree, respectively.
Considering a LiDAR frame rate of 100 ms, Fig.~\ref{fig:CPU_time} reports that only the HSC framework can encode and decode within the LiDAR inter-frame time even with the highest quantization level.

Fig.~\ref{fig:encoding} also exemplifies that, in HSC-1 and HSC-2, the \mbox{RangeNet++} inference time must be considered in addition to that of Draco, namely 56 ms on average using the hardware specified in Sec.~\ref{ssec:results}. 
This preprocessing time could be cut down if considering half of the range image resolution for RangeNet++ (i.e., $[64 \times 1024$]), at the cost of reducing the \gls{iou}, which is used to measure the object detection accuracy in a particular scene. The encoding time can be further decreased if the burden of segmentation is delegated to more powerful cloud computing facilities before transmission.
On the contrary, decoding operations should be executed on-board to ensure real-time responses, thereby making the proposed HSC-1 and HSC-2 schemes (which are faster than HSC-0 and G-PCC) more desirable, as shown in Fig.~\ref{fig:decoding}.
\smallskip
    

\begin{figure}[t!]
        \centering
        \setlength{\belowcaptionskip}{-0.18cm}
            \includegraphics[width=0.88\columnwidth]{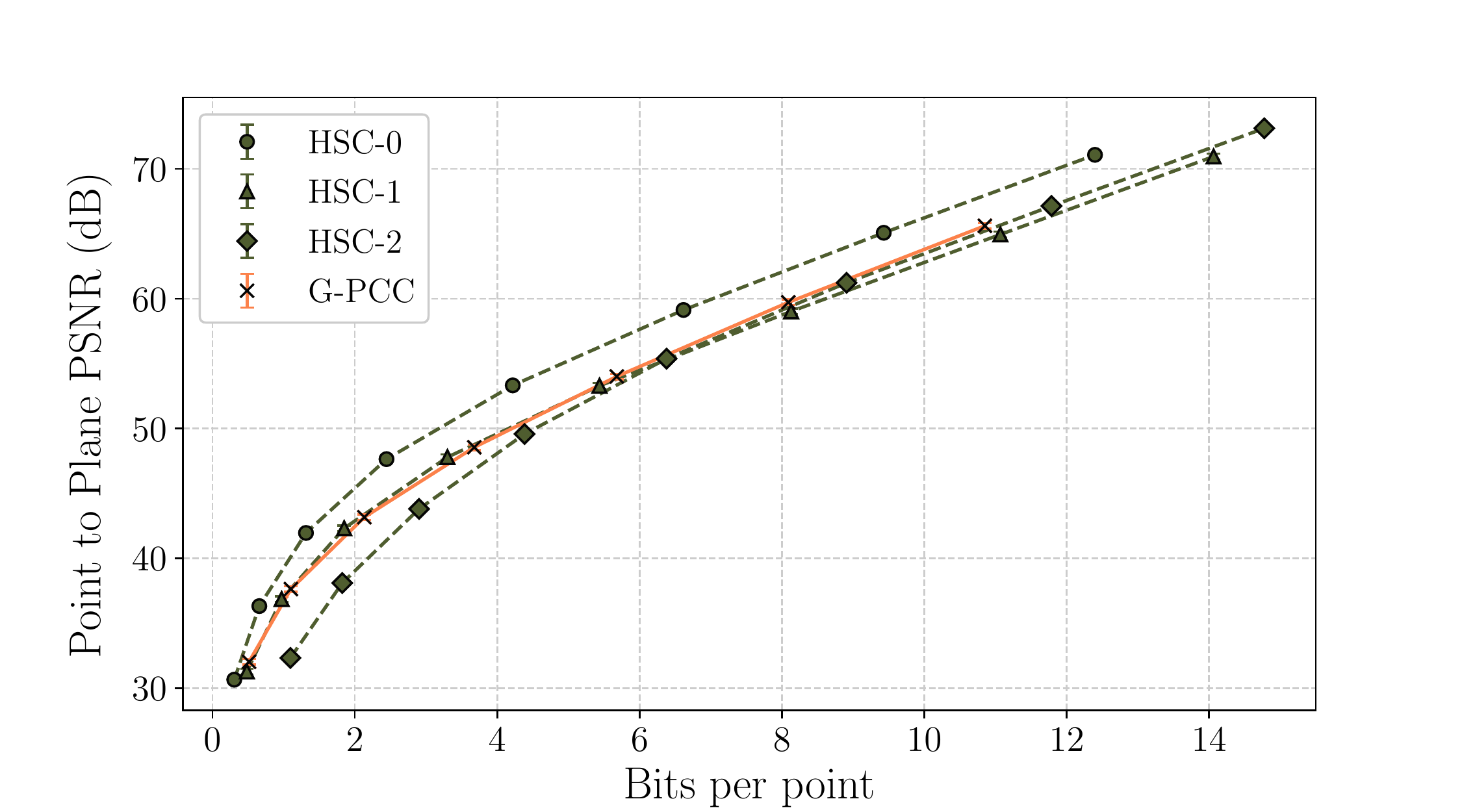}
            \caption{PSNR of the investigated compression strategies vs. the BPP.}%
            \label{fig:psnr}
        \end{figure}

        \begin{figure}[t!]
            \centering
            \includegraphics[width=0.88\columnwidth]{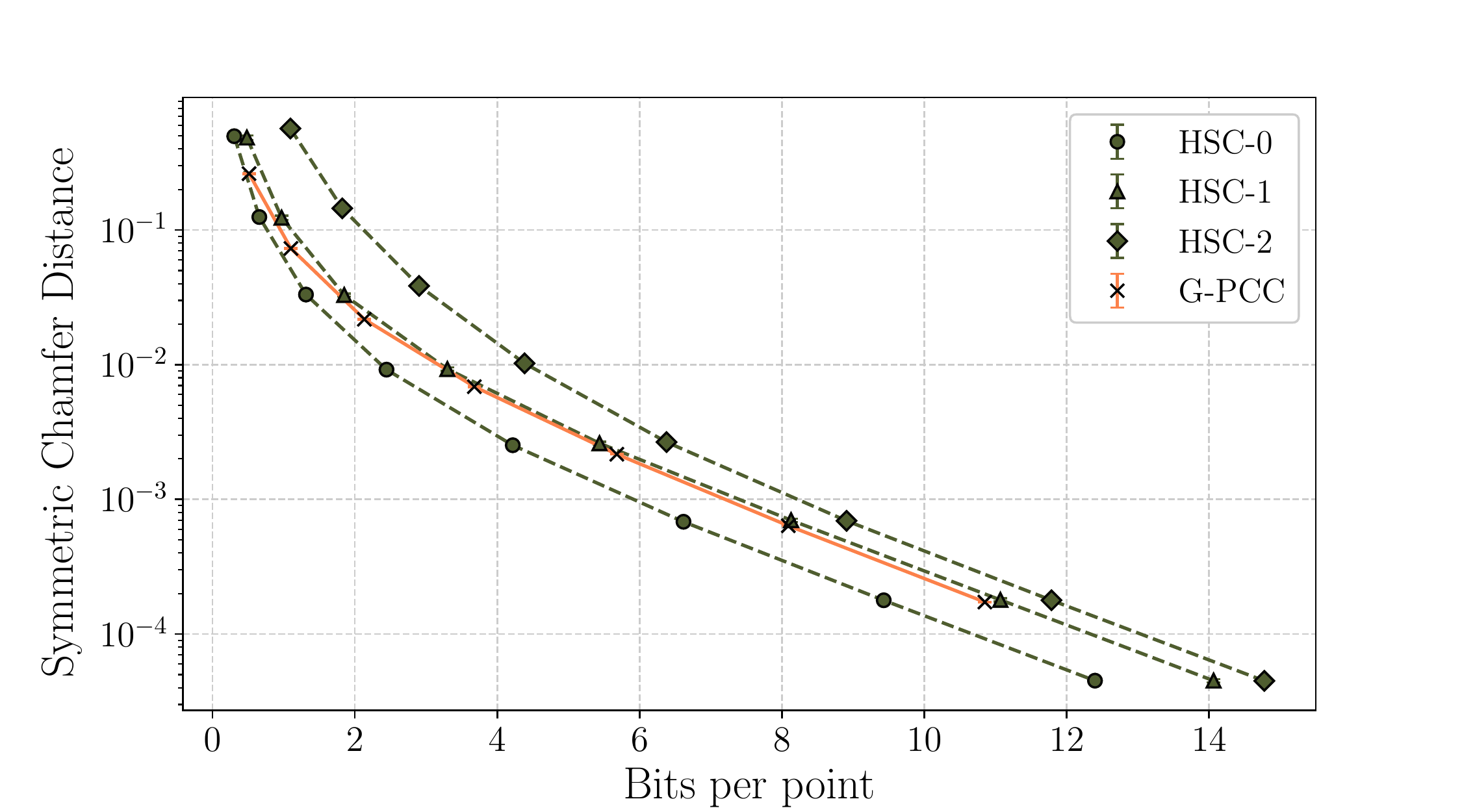}
            \caption{Symmetric point-to-point Chamfer Distance of the investigated compression strategies vs. the BPP.}%
            \label{fig:chamfer}
        \end{figure}

\textbf{Compression accuracy.}
Looking at the PSNR in Fig.~\ref{fig:psnr}, HSC-0 is able to reach the same accuracy of G-PCC with fewer BPP.
In turn, HSC-1 and HSC-2 privilege efficiency over accuracy and result in lower PSNR compared to the baselines. 
Anyway, PSNR reduction is contained to a tolerable $-10\%$ for the most aggressive quantization levels, in the face of an impressive improvement in the file size. 


From Fig.~\ref{fig:chamfer} we also notice  that both HSC and G-PCC methods can perform well in terms of symmetric Chamfer Distance when  limiting the quantization level and the voxelization scaling factor, respectively.

\begin{figure}[t!]
 \center
  \includegraphics[width=\columnwidth]{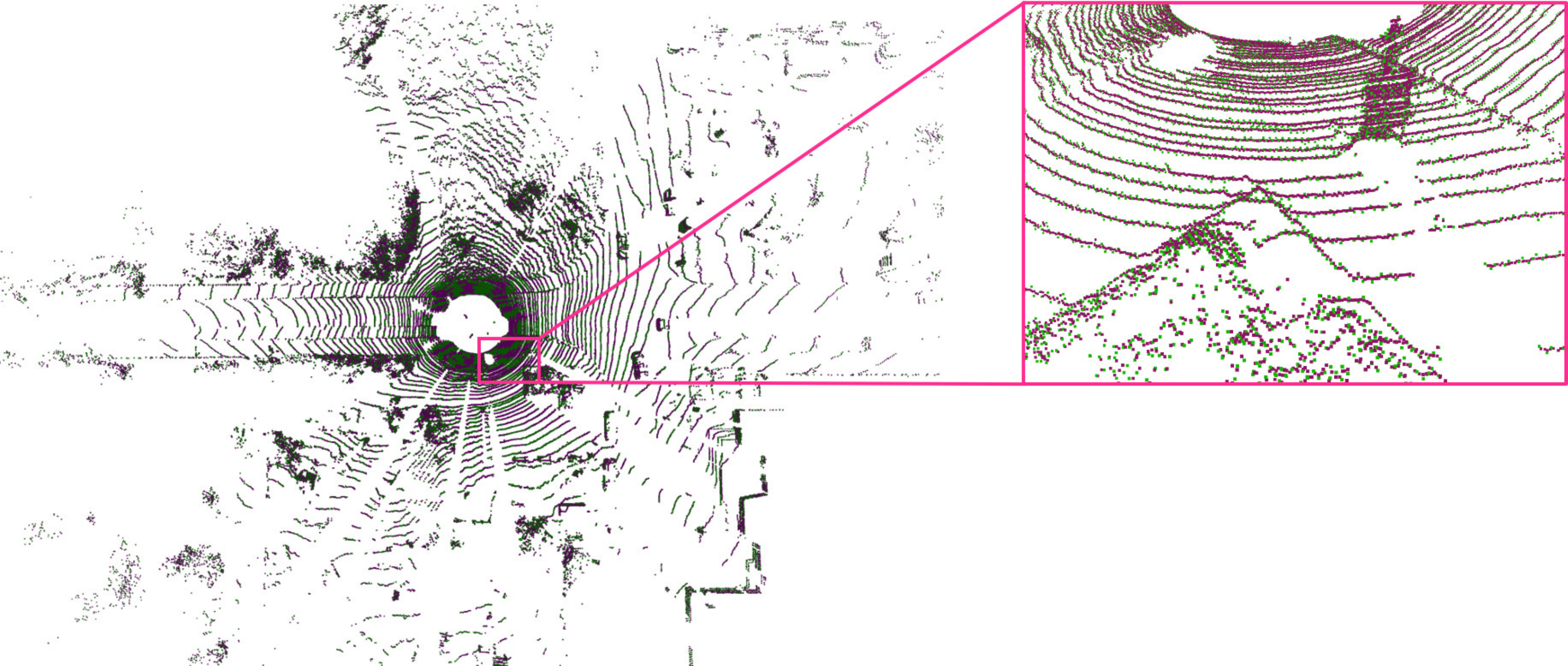}
  \caption{Visual reconstruction of the point cloud obtained with HSC (green) vs. the raw LiDAR acquisition (violet) with quantization level equal to 10.}
  \label{fig:accuracy}
\end{figure}

Based on the previous results, we can conclude that the optimal trade-off between compression efficiency and accuracy in the reconstructed point cloud can be obtained with a quantization level of 9 or 10: in this range, the adoption of semantic segmentation in HSC-2 can achieve up to $8\times$ better compression than using G-PCC.
In practice, inferring data labels and removing the least critical objects from the LiDAR point cloud before transmission guarantee quasi real-time operations as encoding and decoding times never exceed the LiDAR inter-frame time.
Finally, even in the most aggressive, space-saving configuration (i.e., HSC-2), the PSNR is still guaranteed to be above 50 dB (a typical acceptable value for wireless transmission quality loss at which distortions in compressed frames can be hardly noticed~\cite{abramova2014required}) when {BPP $>5$}.
This is also validated in Fig.~\ref{fig:accuracy}, which  shows how the reconstructed point cloud with quantization level equal to 10 is a good approximation of the original LiDAR acquisition.

\section{Concluding Remarks}
\label{sec:conclusions}

In this work we presented an innovative semantic compression algorithm for LiDAR point clouds, designed to achieve quasi real-time transmissions in autonomous driving scenarios.
Our first contribution has been the interesting comparison between G-PCC and Draco for compression, which revealed how the former is not suitable for low-latency and real-time applications.
We then proposed to enhance the compression performance by assisting Draco with \mbox{RangeNet++} semantic segmentation as an approach to preemptively identify the least critical points and defer/cancel their transmissions in favor of more valuable data. 
We validated our framework  on the Semantic KITTI dataset, and we showed that, even though there exists a trade-off between compression efficiency and accuracy, it is possible to compress the raw point cloud by up to 700 times to achieve real-time transmissions with tolerable accuracy degradation.

Future works will consider end-to-end system-level simulations to assess the benefits of the proposed framework in terms of network-related metrics such as the channel occupancy and the overall transmission latency.

\bibliographystyle{IEEEtran}
\bibliography{./biblio}

\end{document}